\newif\ifsubmit
\submitfalse
\ifsubmit
    \documentclass[a4paper,UKenglish,cleveref, autoref, thm-restate]{lipics-v2021}
    \pdfoutput=1 %uncomment to ensure pdflatex processing (mandatatory e.g. to submit to arXiv)
\else
    \documentclass[12pt,hidelinks]{article}
\fi

\ifsubmit\else
    \usepackage[utf8]{inputenc}
    \usepackage{fullpage}
    \usepackage[small,bf]{caption}
    \usepackage{cite}
\fi

\usepackage{amsmath, amsthm, amssymb, url, graphicx, float}
\usepackage{hyperref}
\usepackage{tikz}
\usepackage{pgfplots}

\newcommand{\E}{\mathbf{E}}

\newcommand{\eps}{\varepsilon}

\let\phi\varphi

\newcommand{\ones}{\mathbf 1}
\newcommand{\reals}{{\textbf{\bf R}}}

\newcommand{\conv}{\mathop{\bf conv}}

\DeclareMathOperator*{\argmin}{argmin}

\newcommand{\dom}{\mathop{\bf dom}} % domain

\newcommand{\cf}{{\it cf.}}
\newcommand{\eg}{{\it e.g.}}
\newcommand{\ie}{{\it i.e.}}

\newcommand{\BEAS}{\begin{eqnarray*}}
\newcommand{\EEAS}{\end{eqnarray*}}
\newcommand{\BEA}{\begin{eqnarray}}
\newcommand{\EEA}{\end{eqnarray}}
\newcommand{\BEQ}{\begin{equation}}
\newcommand{\EEQ}{\end{equation}}
\newcommand{\BIT}{\begin{itemize}}
\newcommand{\EIT}{\end{itemize}}

\newif\iftodos

\title{Multidimensional Blockchain Fees are (Essentially) Optimal}
\ifsubmit
    \author{Anonymized}{Anonymized}{anonymous@anon.org}{}{}%TODO mandatory, please use full name; only 1 author per \author macro; first two parameters are mandatory, other parameters can be empty. Please provide at least the name of the affiliation and the country. The full address is optional. Use additional curly braces to indicate the correct name splitting when the last name consists of multiple name parts.

    \authorrunning{Anon, et al.} %TODO mandatory. First: Use abbreviated first/middle names. Second (only in severe cases): Use first author plus 'et al.'

    \Copyright{Anonymized} %TODO mandatory, please use full first names. LIPIcs license is "CC-BY";  http://creativecommons.org/licenses/by/3.0/
    
    
     %TODO mandatory; please add comma-separated list of keywords
    
    \category{} %optional, e.g. invited paper
    
    \relatedversion{} %optional, e.g. full version hosted on arXiv, HAL, or other respository/website
    \acknowledgements{}%optional
    
\else
    \author{Guillermo Angeris\thanks{The authors are listed in alphabetical order}
    \\ \texttt{\small gangeris@baincapital.com} \and Theo Diamandis \\
    \texttt{\small tdiamand@mit.edu} \and Ciamac Moallemi \\ \texttt{\small
    ciamac@gsb.columbia.edu}}
    \date{February 2024}
\fi

\ifsubmit
    \EventEditors{John Q. Open and Joan R. Access}
    \EventNoEds{2}
    \EventLongTitle{42nd Conference on Very Important Topics (CVIT 2016)}
    \EventShortTitle{CVIT 2016}
    \EventAcronym{CVIT}
    \EventYear{2016}
    \EventDate{December 24--27, 2016}
    \EventLocation{Little Whinging, United Kingdom}
    \EventLogo{}
    \SeriesVolume{42}
    \ArticleNo{23}
\fi

\begin{document} 
\maketitle 

\begin{abstract}
    In this paper we show that, using only mild assumptions, previously
    proposed multidimensional blockchain fee markets are essentially optimal,
    even against worst-case adversaries. In particular, we show that the
    average welfare gap between the following two scenarios is at most
    $O(1/\sqrt{T})$, where $T$ is the length of the time horizon considered. In
    the first scenario, the designer knows all future actions by users and is
    allowed to fix the optimal prices of resources ahead of time, based on the
    designer's oracular knowledge of those actions. In the second, the prices
    are updated by a very simple algorithm that does not have this oracular
    knowledge, a special case of which is similar to EIP-1559, the base fee
    mechanism used by the Ethereum blockchain. Roughly speaking, this means
    that, on average, over a reasonable timescale, there is no difference in
    welfare between `correctly' fixing the prices, with oracular knowledge of
    the future, when compared to the proposed algorithm. We show a matching
    lower bound of $\Omega(1/\sqrt{T})$ for any implementable algorithm and
    also separately consider the case where the adversary is known to be
    stochastic.
\end{abstract}

\section{Introduction}
Public blockchains allow any user to submit a transaction that modifies the 
shared state of the network. These transactions are independently verified and 
executed by a decentralized network of full nodes, who must each re-execute the
transaction to verify the correct updated state of the blockchain. Each 
transaction requires some amount of \emph{resources}, such as compute or storage, 
to be executed. Because full nodes have finite resources, blockchains must limit 
the total computational resources that can be consumed per unit of time, usually 
measured per-block. As user demand may fluctuate, most blockchains implement a 
transaction fee mechanism in order to allocate finite resources among competing 
transactions. 

\ifsubmit
    \subparagraph*{Transaction fees.}
\else
    \paragraph{Transaction fees.}
\fi
Transaction fee mechanisms in practice often use a combination of a \emph{base
fee}, which is the minimum cost of entry for a particular block, and a 
\emph{priority fee}, which is an additional fee that a user may pay so that a
particular contentious transaction is included. In this paper, we concentrate
on the base fee for transaction inclusion. These fees are computed algorithmically
as part of the blockchain's consensus mechanism; any transaction not paying the
computed base fee is invalid. Thus, the algorithm for computing these base fees
is a crucial part of the blockchain's design, as it ultimately determines which 
transactions can be included in the blockchain and the blockchain's robustness 
to adversarial behavior, such as spam or denial-of-service attacks.

\ifsubmit
    \subparagraph*{Dynamic base fees.}
\else
    \paragraph{Dynamic base fees.}
\fi
Because user demand fluctuates, many blockchains have implemented or proposed a 
dynamic base fee mechanism. These mechanisms update the base fee based on current 
demand for resources, measured by the resource consumption of the transactions 
in previous blocks. Most famously, EIP-1559~\cite{eip1559} for the Ethereum 
blockchain introduced a base fee that adjusts to track a target block usage.
Many of these mechanisms can be interpreted as applying gradient descent on the
dual of a particular optimization problem~\cite{diamandis2023designing}.

\ifsubmit
    \subparagraph*{Multidimensional fees.}
\else
    \paragraph{Multidimensional fees.}
\fi
Calculating transaction fees through a single, joint unit of account fixes the
relative prices of all resources. These fixed relative prices inhibit granular
price discovery and, ultimately, reduce the throughput of the blockchain. To
mitigate this behavior, some blockchains have proposed \emph{multidimensional
fees}, where different resources are priced separately.
EIP-4844~\cite{buterin2021eip4488}, to be implemented in the upcoming Dencun
update, allows users to submit special pieces of data called
blobs~\cite{adler2019eip2242,adler2019dataavail}, which have an independent fee
mechanism. In essence, EIP-4844 creates a two-dimensional fee market.
SIMD-110~\cite{simd110} in the Solana ecosystem proposes a local fee controller
for each piece of state on the Solana virtual machine. During periods of high
user demand, highly contested pieces of state would become more expensive
without affecting the price of other state. Similar multidimensional mechanisms
have been proposed for the Ethereum
ecosystem~\cite{adler2022ethcc,buterin2022multidimensional}
and other blockchains, such as Avalanche~\cite{ogrady2023fees} and Penumbra~\cite{penumbra2023fees},
have also implemented multidimensional blockchain fees. Like dynamic base
fees, these multidimensional mechanisms can be designed and interpreted by
considering the optimization problem that they attempt to solve.

\ifsubmit
    \subparagraph*{Optimality.}
\else
    \paragraph{Optimality.}
\fi
While the design and incentives of particular existing transaction fee 
mechanisms have been extensively studied, the overall design space has not been 
well-characterized. In particular, it is unclear what properties we can hope for 
from a `good' fee mechanism, the tradeoffs present within the space of `good' 
mechanisms, and what `good' even means in this context, where adversarial 
behavior may preclude the existence of `true' market-clearing prices for 
resources. These open questions are important for the many blockchains currently
working on updated fee mechanisms, especially because many traditional results 
do not apply in the adversarial blockchain environment.

\ifsubmit
    \subparagraph*{This paper.}
\else
    \paragraph{This paper.}
\fi
In this paper, we seek to characterize the optimality of the algorithm for setting
multidimensional blockchain fees proposed in~\cite{diamandis2023designing}. To
judge the optimality of this algorithm, we rely on the notion of \emph{regret}
from the online convex optimization literature~\cite{shalev2012online,hazan2016introduction}.
This quantity measures the difference between two scenarios: in the first, we
dynamically update the resource prices as in~\cite[\S3.4]{diamandis2023designing}; in the second, we assume that the
algorithm knows all future transactions and chooses the best fixed vector of
prices given this knowledge. (The second scenario is similar to the current 
implementation of gas in Ethereum, but, of course, the fixed gas prices of 
individual resources are not known to be optimal.) We show that the average 
welfare gap between the two scenarios is at most $O(1/\sqrt{T})$, where $T$ is 
the length of the time horizon considered, and we construct a corresponding 
lower bound. Taken together our results show that the proposed algorithm
is essentially optimal according to this metric, even under arbitrary 
adversarial behavior that eludes traditional game theoretic analysis. 
We also show that, if the adversary is stochastic, the prices do converge to a
market clearing price and give an explicit rate of convergence. We point to
several additional directions for future work, where we suspect additional
tools from online optimization may be used to improve upon the results
presented here.

\subsection{Related work} 
There are a number of papers proposing new mechanisms for pricing resources or
analyzing mechanisms currently implemented in practice; we review some of the
relevant literature below.

\ifsubmit
    \subparagraph*{Blockchain resource pricing.}
\else
    \paragraph{Blockchain resource pricing.}
\fi
In the specific case of blockchain
resource pricing, these papers can be roughly broken down into three buckets,
which we describe next. The first set analyzes past or current fee mechanisms
or variations thereof. These papers include some of the original economic
analyses of EIP-1559~\cite{roughgarden2021transaction}, empirical
studies~\cite{buterin2018blockchain,liu2022empirical}, and clean theoretical
models~\cite{leonardos2023optimality}. All of these papers focus on the
single-dimensional resource case, usually under some simplifying assumptions
such as, \eg, certain stochastic limits in the case
of~\cite{leonardos2023optimality}. The second set of papers proposes simple
extensions or characterizations of mechanisms for pricing single resources in a
variety of
settings~\cite{ferreira2021dynamic,bahrani2023transaction,basu2023stablefees},
including some impossibility results~\cite{chung2023foundations}. The final set
proposes multidimensional fee mechanisms for a variety of
settings~\cite{buterin2022multidimensional,diamandis2023designing,crapis2023optimal}.
Perhaps the closest in spirit to this particular work
is~\cite{crapis2023optimal}, which proposes a stochastic demand model of
resources with stochastic observations. They apply linear-quadratic control
theory to set prices that minimize a composite loss function that seeks to
balance matching the target block size, while smoothly changing prices. This
work, on the other hand, shows that, even in an adversarial setting, a very
simple update rule suffices and has low regret relative to optimally setting
the prices. We also show that, in the `easier' stochastic setting, we recover a
market-clearing price and provide a rate of convergence.

\ifsubmit
    \subparagraph*{Tatonnement processes.}
\else
    \paragraph{Tatonnement processes.}
\fi
The design of transaction fee mechanisms via
gradient descent on an appropriately-chosen dual problem is also related to the
literature on tatonnement processes~\cite{uzawa1960walras,cole2008fast}. Many
works (\eg,~\cite{bubeck2019multi} and~\cite{roth2020multidimensional}) use
gradient descent-like algorithms to develop dynamic pricing strategies for
various types of markets. These algorithms use a dynamic step size to ensure
convergence to optimal prices. More similar to our
work,~\cite{ashlagi2022price} uses a fixed step size, and their algorithm, as a
result, does not necessarily converge to any single price.

\section{Problem set up}\label{sec:setup}
In this section, we briefly introduce the blockchain resource allocation
problem of~\cite{diamandis2023designing}. We review how the dual problem
leads to a notion of resource prices, which, when set correctly, ensure that
block builders include the `optimal' transactions, \ie, those that maximize the
utility of the transaction producers, minus the loss incurred by the blockchain
for including those transactions.

\subsection{The block building problem}\label{sec:block-building-problem}
We will fix some notation that we use for the remainder of the paper in the
paragraphs that follow and describe the problem we are looking to solve.

\ifsubmit
    \subparagraph*{Transactions and resources.}
\else
    \paragraph{Transactions and resources.}
\fi
First, let $t=1, \dots, T$ denote the
current blockheight. At blockheight $t-1$, we assume that there are $n_t$
transactions from which we may pick a subset to be included in block $t$. (This
list of transactions that we are allowed to choose from is often called the
\emph{mempool}.) Each of these $j=1, \dots, n_t$ transactions has some
associated utility $(q_t)_j \in \reals_+$ which is gained if it is included in
the block. Additionally, transaction $j$, if included in the block, will
consume some amount of the $m$ available resources, which we denote as an
$m$-vector $a_j \in \reals^m$. (The $i$th entry of this vector, $(a_j)_i$,
denotes how much of resource $i$ is consumed by transaction $j$.) Finally, the
vector $x_t \in \{0, 1\}^{n_t}$ will denote the transactions that were included
in block $t$: the $j$th entry $(x_t)_j =1$ if transaction $j$ is included in
block $t$ and is 0, otherwise.

\ifsubmit
    \subparagraph*{Total consumption.}
\else
    \paragraph{Total consumption.}
\fi
This set up lets us write the total resource
consumption of all transactions included in block $t$ in a convenient way.
Define the $m$-vector
\[
    y_t = \sum_{j=1}^n a_j (x_t)_j = A_tx_t,
\]
where the matrix $A_t \in \reals^{m \times n_t}_+$ has $j$th column $a_j$
denoting the resources consumed by transaction $j$ in the mempool. We then have
that $(y_t)_i$ denotes the total amount of resource $i$ is consumed by the
transactions included in block $t$.

\ifsubmit
    \subparagraph*{Allowable transactions.}
\else
    \paragraph{Allowable transactions.}
\fi
We denote the set of allowable transactions
in a block by $S_t \subseteq \{0,1\}^{n_t}$, which may include resource limits
(\eg, if $x \in S_t$, then $Ax \le b$), and interactions among transactions.
Examples may include complementarity constraints such as, if $x \in S_t$, then
$x_j = 1$ implies that $x_{j'} = 0$ for some $j' \neq j$, among many other,
potentially very complicated, constraints.

\ifsubmit
    \subparagraph*{Loss function.}
\else
    \paragraph{Loss function.}
\fi
Finally, we denote the network's `unhappiness' with
the resource usage at block $t$ as the loss function $\ell_t: \reals^m_+ \to
\reals \cup \{\infty\}$. This function maps the resource usage at block $t$,
which is always a nonnegative vector, to some number $\ell_t(y_t)$. Higher
values denote less desirable resource usage. We assume this function, which is
chosen by the blockchain designer, is convex and lower semicontinuous for the
remainder of the paper.

\ifsubmit
    \subparagraph*{Block building problem.}
\else
    \paragraph{Block building problem.}
\fi
We now define the \emph{block building
problem} as the problem of optimizing the utility of the included transactions,
minus the loss incurred by the network, subject to the transactions being
in the allowable transaction set:
\begin{equation}
    \label{eq:resource-allocation}
    \begin{aligned}
        & \text{maximize} && q^T_tx - \ell_t(y_t)\\
        & \text{subject to} && y_t = A_tx_t  \\
        &&& x_t \in \conv(S_t).
    \end{aligned}
\end{equation}
Here, the variables are the included transactions $x_t \in \reals^{n_t}$ and the
total resource consumption of the included transactions, $y_t \in \reals^{m}$.
Additionally, $\conv(S_t)$ denotes the convex hull of the set $S_t$. Replacing $S_t$
with its convex hull allows `fractional' transactions, such as $x_j = 1/3$.
These values of $x_j \in (0,1)$ can be interpreted as `spreading out'
transaction $j$ over the next $1/x_j$ blocks. We denote the optimal value of
problem~\eqref{eq:resource-allocation} by $s^\star$.

\ifsubmit
    \subparagraph*{Bounds.}
\else
    \paragraph{Bounds.}
\fi
In general, it is a good idea to have maximum per-block
resource consumption bounds, which we write as $b \in \reals^m_+$, so we will
assume that
\[
    S_t \subseteq \{x \in \{0, 1\}^{n_t} \mid A_tx \le b\},
\]
so any allowable choice of transactions $x_t \in S_t$ will satisfy
\[
    0 \le A_tx_t \le b,
\]
as both $A_t$ and $x_t$ are nonnegative. In other words, we assume that the
maximum amount of resource $i$ that any set of acceptable transactions may
consume is no more than $b_i$. Note that this also implies
\[
    0 \le y_t \le b,
\]
for any $y_t$ feasible for~\eqref{eq:resource-allocation} since $y_t = A_tx_t$.
(For why such limits are necessary, see the discussion
in~\cite[\S1]{diamandis2023designing}.) Since $0 \le y_t \le b$ for any
feasible points, we assume, without loss of generality, that $\ell_t(y) =
\infty$ whenever $y \not \le b$ or $y \not \ge 0$ for convenience.

\ifsubmit
    \subparagraph*{Discussion.}
\else
    \paragraph{Discussion.}
\fi
We may view this block building problem as the
following `idealized' scenario: if we, as the network, knew the welfare
generated by each transaction, along with all possible constraints for all
transactions then solving problem~\eqref{eq:resource-allocation} would give us
a way to include transactions as to maximize welfare minus the loss incurred by
the network. Of course, in general, almost all of these things are not known
(or, potentially even knowable) to the network, so we will show how to
approximately solve this problem by, instead, correctly setting the price of
the consumed resources for each block.

\subsection{The dual problem}
In this subsection, we construct a particular dual problem, whose variables are
the prices of the $m$ possible resources; correctly setting these prices would
solve our original problem~\eqref{eq:resource-allocation}. We then show that
its gradients can be efficiently evaluated, without knowledge of things like
the allowable transactions $S_t$ nor the welfare $q_t$ for each block $t$. This
suggests a heuristic in which we use the gradient at each step to update the
prices of the resources, which we prove in a later section is (essentially)
optimal.

\ifsubmit
    \subparagraph*{The dual function.}
\else
    \paragraph{The dual function.}
\fi
One possible relaxation
of~\eqref{eq:resource-allocation}, for each block $t=1, \dots, T$, can be
constructed by relaxing the equality constraint
in~\eqref{eq:resource-allocation} to a penalty, parametrized by a vector $p_t
\in \reals^m$, which we will call the \emph{prices}:
\[
    \begin{aligned}
        & \text{maximize} && q^T_tx - \ell_t(y_t) + p_t^T(y_t - A_tx_t)\\
        & \text{subject to} && x_t \in \conv(S_t),
    \end{aligned}
\]
where the variables are the same as those of
problem~\eqref{eq:resource-allocation}. If we denote the optimal value of this
problem, which is parametrized by the vector $p_t$, as $f_t(p_t)$, then, since
the objective is separable over $x_t$ and $y_t$, we can write the optimal value
of this problem as
\begin{equation}\label{eq:dual-fn}
    f_t(p_t) = \sup_{y} \left(p^T_ty - \ell_t(y)\right) + \sup_{x \in \conv(S_t)} (q_t - A_t^Tp_t)^Tx,
\end{equation}
We note that this function $f_t$, which we will call the \emph{dual function},
is convex since it is the pointwise supremum of a family of functions linear in
$p_t$.

\ifsubmit
    \subparagraph*{Fenchel conjugate.}
\else
    \paragraph{Fenchel conjugate.}
\fi
The first term of~\eqref{eq:dual-fn} can be
recognized as the Fenchel conjugate of $\ell$~\cite[\S3.3]{cvxbook} evaluated
at $p_t$, which we write as $\ell^*_t(p_t)$. This function can be interpreted
as choosing the resource allocation $y$ that maximizes the network's profit
minus the loss as a result of resource consumption $y$. We note that the
Fenchel conjugate has a closed-form expression for many commonly used loss
functions.

\ifsubmit
    \subparagraph*{Block building problem.}
\else
    \paragraph{Block building problem.}
\fi
The second term in~\eqref{eq:dual-fn} is
the optimal value of the following problem:
\[
    \begin{aligned}
        & \text{maximize} && (q_t - A^T_t p_t)^Tx\\
        & \text{subject to} &&x \in \conv(S_t),
    \end{aligned}    
\]
with variable $x \in \reals^n$. We can interpret this problem as the
transaction producers' problem of creating and choosing transactions to be
included in a block in order to maximize the utility of the included
transactions ($q_t^Tx$), after netting the fees paid to the network ($(A^T_t
p_t)^Tx$). (This problem is sometimes called the \emph{block building problem}
or \emph{packing problem} that is solved by block producers. These `block
producers' include both users and validators.) Since the objective of
this problem is linear, it has the same optimal value as the nonconvex problem
\begin{equation}\label{eq:welfare-max}
    \begin{aligned}
        & \text{maximize} && (q_t - A^T_t p_t)^Tx\\
        & \text{subject to} &&x \in S_t,
    \end{aligned}
\end{equation} 
with variable $x \in \reals^n$. Note that the packing problem problem depends
on the problem data associated with block $t$: the utilities $q_t$, the 
resources required by the submitted transactions, $A_t$, and the associated 
constraints $S_t$ for a particular block at time $t$. We will denote the optimal 
value of the packing problem as $h_t(p)$, where the index $t$ indicates that the 
problem data is associated with block $t$.

\ifsubmit
    \subparagraph*{Gradient.}
\else
    \paragraph{Gradient.}
\fi
With this notation, the dual function can be written
as 
\[
    f_t(p) = \ell^*_t(p) + h_t(p).
\]
We note, for future use, that, when $f_t$ is differentiable at $p$,
\begin{equation}\label{eq:grad}
    \nabla f_t(p) = y^\star_t(p) - A_tx^\star_t(p),
\end{equation}
where $y^\star_t(p)$ denotes the maximizer for the Fenchel conjugate of
$\ell_t$, at price $p$, while $x^\star_t(p)$ denotes the optimal transactions
to be included for the block building problem~\eqref{eq:welfare-max} at price
$p$. If $f_t$ is not differentiable at $p$, then any solution to each
subproblem suffices. Note that both of these quantities are easy to compute if
price $p$ is set at blockheight $t-1$ and block $t$ is submitted with these
resource prices $p$: the former term, $y^\star(p)$, is the solution to a basic
problem that often has a closed form, while the latter term, $Ax^\star(p)$, is
exactly the total resource consumption of the transactions included in block
$t$.

\ifsubmit
    \subparagraph*{Bounds on the gradient.}
\else
    \paragraph{Bounds on the gradient.}
\fi
From the discussion in~\S\ref{sec:block-building-problem}, 
we have that the allowable transactions in $S_t$ can only consume up to some 
resource limit $b$, which means that:
\[
    0 \le A_tx_t^\star(p) \le b,
\]
from problem~\eqref{eq:welfare-max}. Additionally, since $\ell_t(y) = \infty$
unless $0 \le y \le b$, as noted in~\S\ref{sec:block-building-problem}, then, 
it is not hard to show that
\[
    0 \le y^\star_t(p) \le b,
\]
so we receive the simple bound on the gradient of $f_t$ at $p$:
\[
    -b \le y_t^\star(p) - A_tx^\star_t(p) \le b.
\]
This implies the gradient bound
\[
    \|\nabla f_t(p)\|_s \le \|b\|_s
\]
for any $s$-norm.

\subsection{The aggregated dual problem}\label{sec:dual-problem}
If the dual function $f_t$ is differentiable, then, since it is convex, finding a
minimizer of the dual function corresponds exactly to finding a point $p^\star$
satisfying
\[
    \nabla f_t(p^\star) = 0.
\]
Using~\eqref{eq:grad}, this is the same as finding a set of prices $p^\star$
such that
\[
    A_tx^\star_t(p^\star) = y^\star_t(p^\star).
\]
We can interpret the left term as the demand for resources at price $p^\star$
(from users and nodes) and the right term as the supply of resources at price
$p^\star$ (from the network). Setting the prices correctly balances the demand
and supply of the available resources. One can also show that, when $f_t$ is
differentiable and this condition is satisfied, the transactions
$x^\star_t(p^\star)$ and $y^\star_t(p^\star)$ exactly solve the original
problem~\eqref{eq:resource-allocation}.

\ifsubmit
    \subparagraph*{The (aggregated) dual problem.}
\else
    \paragraph{The (aggregated) dual problem.}
\fi
As blockchain designers, we would
like to set prices that, at least on average, render the supply and demand of
resources (approximately) equal. That is, we would like to find a set of prices
that solves the problem:
\begin{equation}\label{eq:agg-dual}
    \begin{aligned}
        & \text{minimize} && \sum_{t=1}^T f_t(p),
    \end{aligned}
\end{equation}
over $p \in \reals^m$. Finding a price $p^\star$ that minimizes this objective means
that, using~\eqref{eq:resource-allocation}, over the $T$ blocks, we have
\[
    \sum_{t=1}^T A_tx_t^\star(p^\star) = \sum_{t=1}^T y_t^\star(p^\star),
\]
\ie, that the demand and supply for resources over the $T$ blocks are both
equal. Or, in other words, that the market, over all $T$ blocks, clears on
aggregate when the prices $p$ are correctly set. (The case where the objective
is not differentiable over $p^\star$, but is subdifferentiable, means that
there is at least one solution to each subproblem, at optimal prices $p^\star$,
for which supply and demand are both equal.)

\ifsubmit
    \subparagraph*{A basic heuristic.}
\else
    \paragraph{A basic heuristic.}
\fi
Since the prices $p$ for the blockchain must be
set before the blocks are observed, it would be hard to
solve~\eqref{eq:agg-dual} without oracular knowledge of the functions $f_t$.
A simple heuristic is then the following: start with some prices
$p_t$, set at blockheight $t-1$, observe what happens at block $t$ with these
prices, and update the prices slightly (by, say $\eta > 0$) according to the
gradient information:
\begin{equation}\label{eq:grad-update}
    p_{t+1} = p_t - \eta \nabla f_t(p_t).
\end{equation}
(If $f_t$ is not differentiable at $p_t$, then replace the gradient with a
subgradient.) From before, we know that the gradient at block $t$ is observable
at the prices set in the previous block, $p_t$. This heuristic has roughly the
`right behavior': if our prices are too low for a certain resource, say $i$,
then the demand at these prices exceeds the expected supply, so
\[
    (y_t(p_t))_i > (A_tx_t^\star(p_t))_i.
\]
From the update step, this would increase the price $(p_t)_i$ of resource
$i$ proportional to this amount. (The opposite is true if the prices at time
$t-1$ were set too high.) See~\cite{diamandis2023designing} for additional 
discussion.

\ifsubmit
    \subparagraph*{A multiplicative update.}
\else
    \paragraph{A multiplicative update.}
\fi
Of course, we can also imagine any number
of other update rules. For example, we may use a multiplicative update:
\begin{equation}\label{eq:mult-update}
    p_{t+1} = p_t \circ \exp(-\eta \nabla f_t(p_t)),
\end{equation}
where $\circ$ denotes the elementwise (Hadamard) product. This update ensures
that the prices are always nonnegative; though the previous
update~\eqref{eq:grad-update} can be modified slightly to accommodate this as well. 
We note that this update rule, when there is one resource, corresponds exactly
to EIP-1559 as implemented as in the EIP-4844 update Ethereum today.

\ifsubmit
    \subparagraph*{Discussion.}
\else
    \paragraph{Discussion.}
\fi
Of course, the next natural question is: just how well
do these heuristics perform? We will show in the remainder of the paper that
the average gap between the objective values provided by this heuristic and
those given by setting the optimal prices $p^\star$ (which clear the market on
average; \ie, those that solve problem~\eqref{eq:agg-dual}) tends to zero as
the number of blocks in question, $T$, grows:
\begin{equation}\label{eq:avg-reg-inequality}
    \frac{1}{T}\left(\sum_{t=1}^T f_t(p_t) - \sum_{t=1}^T f_t(p^\star)\right) \le \frac{C}{\sqrt{T}},
\end{equation}
for any (reasonable) choice of $p$. Perhaps the most surprising part is that we
will show that this is true \emph{even when the allowable transactions $S_t$,
resource utilizations $A_t$, and the welfare $q_t$ are controlled by an
adversary}. We will also show that these algorithms are essentially optimal: no
better rate, in terms of $T$, is possible for any algorithm with such an
adversary. The fact that this is possible, along with the original bounds for
such problems, come from a long line of work in online convex
optimization~\cite{shalev2012online}, which we present (and adapt to our specific scenario)
in the next section.

\section{Bounding the gap}
From~\S\ref{sec:setup}, the main quantity we wish to bound is the difference
between the algorithm's performance (\ie, the aggregate performance of the
$p_t$) against the performance of the aggregate market clearing price (\ie,
the price $p^\star$ that minimizes~\eqref{eq:agg-dual}):
\begin{equation}\label{eq:regret-def}
    R = \sum_{i=1}^T f_t(p_t) - \sum_{i=1}^T f_t(p^\star).
\end{equation}
This quantity $R$ is called the \emph{regret} of the algorithm. Bounding this
regret for a variety of algorithms for choosing $p_t$ is the main goal of the
field of online algorithms, for which there are many great resources,
\cf,~\cite{shalev2012online,hazan2016introduction}. A bound on this quantity
$R$ would then also give us a bound of the form of~\eqref{eq:avg-reg-inequality}
given by $R/T$, which is called the \emph{average regret}. We will give a very
short and self-contained exposition of online convex optimization, tailored to
our special case, in the what remains of this section and in the next section.

\ifsubmit
    \subparagraph*{Discussion.}
\else
    \paragraph{Discussion.}
\fi
Note that this comparison is interesting for a few
reasons. First, the price $p^\star$ can only be computed if all of the dual
functions $f_t$ are known in advance, which would require knowledge that the 
network does not have. (On the other hand, updating the prices $p_t$ via a rule 
such as~\eqref{eq:grad-update} is very easy to implement in practice.) The 
second observation is that it need not be true that $p_t$ approaches $p^\star$, 
except in very special cases, even when the regret is low. (One case where this 
is true is, for example, when the $f_t = f$, for some fixed function $f$, which 
is a common simplification in some analyses of dynamic fees, or when the $f_t$ 
are i.i.d.\ drawn from some fixed distribution and strongly convex.)

\ifsubmit
    \subparagraph*{Linearization.}
\else
    \paragraph{Linearization.}
\fi
Since the functions $f_t$ are convex,
then, by definition~\cite[\S3.1.3]{cvxbook} we have that
\[
    f_t(p^\star) \ge f_t(p_t) + \nabla f_t(p_t)^T(p^\star - p_t).
\]
Rearranging, this gives an upper bound on the difference in objective value at
time $t$,
\[
    f_t(p_t) - f_t(p^\star) \le \nabla f_t(p_t)^T(p_t - p^\star).
\]
For the remainder of the paper, we will write $g_t = \nabla f_t(p_t)$ for
notational convenience. (If $f_t$ is not differentiable at $p_t$, then any
subgradient will suffice.) Summing both sides over $t$ then gives
\begin{equation}\label{eq:linear-reg}
    R \le \sum_{i=1}^T g_t^T(p_t - p^\star).
\end{equation}
In other words, if we can find a reasonable upper bound for the linear term on
the right hand side, we can give a bound on the regret $R$, or the average
regret $R/T$.

\ifsubmit
    \subparagraph*{Results.}
\else
    \paragraph{Results.}
\fi
The main point of this section will be to use
bound~\eqref{eq:linear-reg}, along with a simple framework to show that the
linear update rule~\eqref{eq:grad-update}, with appropriately chosen step size
$\eta > 0$ has average regret bounded from above by
\[
    \frac{R}{T} = \frac{1}{T}\left( \sum_{i=1}^T f_t(p_t) - \sum_{i=1}^T f_t(p^\star)\right) \le \frac{BM}{\sqrt{T}},
\]
where $\|b\|_2 \le B$ is an upper bound on the norm of the resource limits,
while $\|p^\star\|_2 \le M$ is a bound on the optimal price. In the case that
the adversary is stochastic (\ie, if the $f_t$ are random functions, uniformly
and identically drawn under certain conditions we state below) there is a much
stronger result. Here, the time-averaged price $\bar p = (1/T)\sum_{t=1}^T p_t$
converges to the optimal value in expectation in that
\[
    \E\left[f(\bar p) - f(p^\star)\right] \le \frac{B^2M}{\sqrt{T}},
\]
using again rule~\eqref{eq:grad-update}, where $f(p) = \E[f_1(p)]$, for any
fixed $p$. (We can, of course, replace $f_1$ with any other $f_t$.) This also
implies that, if the $f_t$ are almost surely $\mu$-strongly convex, for some
$\mu > 0$, then $\bar p$ converges to the optimal price $p^\star$ in
expectation. Finally, if the update used is the multiplicative rule
of~\eqref{eq:mult-update}, then the average regret can be bounded from above by
\[
    \frac{R}{T} \le BM\sqrt{\frac{m\log(M/\eps)}{2T}}.
\]
Here, we define $\|p^\star\|_\infty \le M$, while $B = \|b\|_2$, as before,
while $\eps \le 1$ is a lower bound on the price used in the first iteration of
the rule, $p_0 \ge \eps \ones$, where $\ones$ is the all-ones vector and the
inequality is elementwise.

\subsection{Choice functions and updates}\label{sec:choice-functions}
We will focus on the two heuristics presented in~\S\ref{sec:dual-problem} to
update the prices and show that, for each, the regret is low. To do this, we
rewrite the algorithms as a special case of a general update rule, sometimes
known as FTRL or Follow The Regularized
Leader~\cite{shalev2012online,hazan2016introduction}, which we will then use to
find a bound on the regret. The presentation here is a slight variation of the
lovely theorem and proof of~\cite[\S2]{kakade2009regularization}.

\ifsubmit
    \subparagraph*{Choice function.}
\else
    \paragraph{Choice function.}
\fi
Given some \emph{choice function} $F: \reals^m \to
\reals \cup \{\infty\}$, we choose the new prices $p_{t+1}$ given the previous
gradients $g_1, \dots, g_t$ by setting
\begin{equation}\label{eq:price-def}
    p_{t+1} = \nabla F(-\eta(g_1+\dots +g_t)),
\end{equation}
where $\eta > 0$ will be some specific step size to be chosen later. (The
gradient can be replaced with any subgradient whenever the function is not
differentiable, but is convex.) The one requirement we will have of this
function $F$ is that it is `smooth'; in particular that the Hessian of $F$ (if
twice-differentiable) is bounded from above by $\sigma I$, where $\sigma > 0$,
in some suitable domain $D \subseteq \reals^n$.

\ifsubmit
    \subparagraph*{Gradient descent.}
\else
    \paragraph{Gradient descent.}
\fi
One choice of functions we will
use is the norm-squared,
\[
    F(z) = \frac12 \|p_0 + z\|_2^2.
\]
Since $\nabla F(z) = p_0 + z$, then
\[
    p_{t+1} = p_0 -\eta (g_1 + \dots + g_t) = p_t -\eta g_t,
\]
which corresponds exactly to the gradient descent rule proposed
in~\eqref{eq:grad-update}. This particular choice function is smooth with
$\sigma=1$ for any domain $D$.

\ifsubmit
    \subparagraph*{Multiplicative update.}
\else
    \paragraph{Multiplicative update.}
\fi
The second choice function is
\[
    F(z) = \sum_{i=1}^m (p_0)_i\exp(z_i),
\]
which leads to the multiplicative update rule found in~\eqref{eq:mult-update},
\[
    p_{t+1} = p_0 \circ \exp(-\eta(g_1+\dots + g_t)) = p_t \circ \exp(-\eta g_t).
\]
Here $\exp(\cdot)$ is taken elementwise, while $\circ$ denotes the elementwise
(Hadamard) product. For the domain $D = \{z \mid \|z\|_\infty \le \log(M)\}$,
this choice function is smooth with $\sigma = \|p_0\|_\infty M$. (We use
$\log(M)$ since a bound on the prices of order $M$ would imply a bound on the
domain size of order around $\log(M)$---this is a somewhat technical condition
that can be avoided by a slightly more careful choice of $F$;
\cf,~appendix~\ref{app:alt-mult-choice}.) We provide an alternative 
interpretation of this rule in appendix~\ref{app:eip-1559}.

\subsection{Constructing general regret bounds}
We will combine two simple observations to give a bound on
the linear term of~\eqref{eq:linear-reg}. The first is that, if we define
the Fenchel conjugate of $F$ over the domain $D$ by
\[
    F^*(\tilde p) = \sup_{z \in D} (\tilde p^Tz - F(z)),
\]
then we have the definitional inequality
\begin{equation}\label{eq:conj}
    z^T\tilde p - F^*(\tilde p) \le F(z),
\end{equation}
for any choice of $\tilde p \in \reals^m$ and $z \in D$. The second is that,
since $F$ is smooth with constant $\sigma$, we have, by a simple
exercise in integration:
\[
    F(z + z') \le F(z) + \nabla F(z)^Tz' + \frac{\sigma}{2}\|z'\|_2^2.
\]
So, given a sequence of $g_1, \dots, g_T$, by applying this inequality
repeatedly and using the definition of the prices $p_t$ given
in~\eqref{eq:price-def}, we have
\[
    F(-\eta(g_1 + \dots + g_T)) \le F(0) - \eta\sum_{t=1}^T g_t^Tp_t + \frac{\sigma\eta^2}{2}\sum_{t=1}^T \|g_t\|_2^2.
\]
Finally, setting $z=-\eta(g_1+\dots + g_T)$ in~\eqref{eq:conj} and using the above
inequality gives
\[
    \sum_{t=1}^T g_t^T(p_t - \tilde p) \le \frac{F(0) + F^*(\tilde p)}{\eta} + \frac{\sigma\eta}{2}\sum_{t=1}^T \|g_t\|_2^2.
\]
Since the gradients $g_t$ are bounded from above by $\|g_t\|_2 \le \|b\|_2 = B$, this
leads to the following bound on the linearized quantity:
\begin{equation}\label{eq:regret-bound}
    \sum_{t=1}^T g_t^T(p_t - \tilde p) \le B\sqrt{\sigma T(F(0)+F^*(\tilde p))},
\end{equation}
by choosing the step size
\[
    \eta = \sqrt{\frac{F(0)+F^*(\tilde p)}{\sigma TB^2/2}},
\]
for any choice of $\tilde p$.

\ifsubmit
    \subparagraph*{Regret bound.}
\else
    \paragraph{Regret bound.}
\fi
Using~\eqref{eq:regret-bound}, we can now give a
bound on the regret $R$ using the inequality of~\eqref{eq:linear-reg}. More
specifically, if we know that the optimal aggregate clearing price $p^\star$
lies in some set $P$, then, using~\eqref{eq:regret-bound}, we have
\[
    R \le B\sqrt{\sigma T(F(0)+F^*(p^\star))} \le B\sqrt{\textstyle\sigma T(F(0)+\sup_{p \in P} F^*(p))},
\]
or, alternatively, that the average regret over that time horizon, $R/T$, is
\begin{equation}\label{eq:avg-regret-bound}
    \frac{R}{T} \le B\sqrt{\frac{\textstyle\sigma (F(0)+\sup_{p \in P} F^*(p))}{T}}.
\end{equation}

\subsection{Regret bounds for updates}\label{sec:reg-bounds-update}
Given the set up above, we are now ready to give bounds on the regret
for both the gradient-descent-like update rule of~\eqref{eq:grad-update}
and the multiplicative update of~\eqref{eq:mult-update}.

\ifsubmit
    \subparagraph*{Norm-squared choice function.}
\else
    \paragraph{Norm-squared choice function.}
\fi
For the gradient-descent-like
rule~\eqref{eq:grad-update}, we have the choice function $F(z) = \frac12 \|p_0
+ z\|^2_2$. Applying the bound~\eqref{eq:regret-bound} gives, since $F(0) =
\|p_0\|_2^2$ and $\sigma = 1$ for this choice function
(from~\S\ref{sec:choice-functions}):
\[
    \frac{1}{T}\sum_{t=1}^T g_t^T(p_t - \tilde p) \le B\sqrt{\frac{\|p_0\|_2^2 + F^*(\tilde p)}{2T}}.
\]
It is not hard to show that $F^*(\tilde p) = \frac12 \|\tilde p\|_2^2$, and,
assuming that we have a bound $\|p^\star\|_2 \le M$ (and similarly for $p_0$),
then the average regret is bounded by~\eqref{eq:avg-regret-bound},
\[
    \frac{R}{T} \le \frac{BM}{\sqrt{T}}.
\]

\ifsubmit
    \subparagraph*{Exponential choice function.}
\else
    \paragraph{Exponential choice function.}
\fi
In the multiplicative update
rule~\eqref{eq:mult-update}, the exponential choice function is given by $F(z)
= \sum_{i=1}^m\exp(z_i)$ has $F(0) = \|p_0\|_1$ and $\sigma = M$ such that,
again, using bound~\eqref{eq:regret-bound},
\begin{equation}\label{eq:regret-mult-update}
    \frac{1}{T}\sum_{t=1}^T g_t^T(p_t - \tilde p) \le B\sqrt{\frac{M(\|p_0\|_1 + F^*(\tilde p))}{2T}}.
\end{equation}
It is also not hard to show that the conjugate of $F$ is
\[
    F^*(\tilde p) = \sum_{i=1}^m \tilde p_i \log \frac{\tilde p_i}{(p_0)_i} - \tilde p_i,
\]
and, if we have a bound that $0 \le \tilde p \le M\ones$ and $\eps \ones \le
p_0 \le M\ones$ for $\eps \le 1$, then
\[
    \sup_{\|\tilde p\|_\infty \le M} F^*(\tilde p) \le m\max\{M\log(M/\eps) - M, 0\},
\]
which follows from the fact that $F^*$ is separable and convex so is maximized
at the boundary of the domain. For simplicity, we will assume that $M \ge e$,
in which case the first term in the max is nonnegative. Putting it all together,
we know that
\[
    \|p_0\|_1 \le mM \quad \text{and} \quad F^*(p^\star) \le m (M\log (M/\eps) - M).
\]
Plugging these values into~\eqref{eq:regret-mult-update} gives
\[
    \frac{R}{T} \le BM\sqrt{\frac{m \log (M/\eps)}{2T}},
\]
where, from before, $M$ is the maximum price allowed for any one resource, $m$
is the number of resources, and $\eps$ is the minimum starting price. The
latter condition cannot be dropped since, if $p_0 = 0$, no update would ever be
able to increase the price, while on the order of $\sim \log(1/\eps)$ updates
are needed to scale the price $p_0$ to any constant. A sufficient condition to
ensure that $p_0 \ge \eps \ones$ is that there is always some minimal demand
for blockspace at a small enough price, as in,
\eg,~\cite[\S3.2]{diamandis2023designing}.

\subsection{A stochastic model}\label{sec:stochastic}
In this subsection, we consider a stochastic setting instead of the adversarial
one above, which will give us stronger results on the convergence of the prices
$p_t$. In particular, given a probability space with sample space $\Omega$ and
distribution $P$, we assume that a sample $\omega_t\in\Omega$ determines the
parameters of the block building problem at time $t$, so that,
\[
q_t = \tilde q(\omega_t), \quad S_t = \tilde S(\omega_t), \quad \ell_t(\cdot) = \tilde \ell(\cdot,\omega_t), \quad A_t = \tilde A(\omega_t),
\]
where the welfare $\tilde q$, feasible set $\tilde S$, losses
$\ell(\cdot,\omega)$ (viewed as a function over $\omega$), and consumption
matrices $\tilde A$ are random variables over $\Omega$. Given a sample $\omega
\in \Omega$ and a price vector $p$, the utility in scenario $\omega$ for price
$p$ is given by the concave function
\[
  \tilde f(p,\omega) = \sup_{y} \left(p^T y - \tilde \ell(y,\omega)\right)
  + \sup_{x \in \conv(\tilde S(\omega))} (\tilde q(\omega) - \tilde A(\omega)^T p)^Tx.
\]
(In other words, we replace the problem data in the definition of the dual
function~\eqref{eq:dual-fn} with the corresponding sampled versions.) Here, we
will show that, in a certain sense, the time-averaged price of the algorithm
minimizes the expected utility,
\begin{equation}\label{eq:expected-util}
   f(p) = \E[\tilde f(p, \omega)].
\end{equation}
For later, we will define $\tilde p^\star \in \argmin f$. Unlike in the
previous set up for adversarial functions, in this problem, the observed
functions are random variables $f_t(p) = \tilde f(p, \omega_t)$, where the
$\omega_t$ are drawn i.i.d.\ according to $P$. So, in this setting, the
gradient update rule of~\eqref{eq:grad-update},
\[
  p_{t+1} = p_t - \eta \nabla f_t(p_t),
\]
corresponds to an instance of stochastic gradient descent with stepsize given
by $\eta > 0$.

\ifsubmit
    \subparagraph*{Result.}
\else
    \paragraph{Result.}
\fi
A standard result in stochastic gradient descent
from, say~\cite[Thm.\ 9.5]{garrigos2023handbook}, is that,
\[
    \E[f(\bar p)] - f(\tilde p^\star) \le \frac{2M^2}{\eta T} + \frac{\eta B^4}{2}.
\]
Here, $B = \|b\|_2$ and $\|\tilde p^\star\|_2 \le M$ is a bound on the optimal
price $\tilde p^\star$ that minimizes the expected utility $f$ given
in~\eqref{eq:expected-util}. (We also assume that the initial price $\|p_0\|_2 \le
M$.) Finally, the time-averaged price $\bar p$ is defined
\[
    \bar p = \frac{1}{T}\sum_{t=1}^T p_t.
\]
Setting $\eta = 2M/B^2\sqrt{T}$ then gives the bound:
\[
    \E[f(\bar p)] - f(\tilde p^\star) \le \frac{B^2M}{\sqrt{T}}.
\]

\ifsubmit
    \subparagraph*{Extensions.}
\else
    \paragraph{Extensions.}
\fi
There is an immediate extension to this result. The
first is to note that, by assumption, this result relies on i.i.d.\ draws of
the sample $\omega_t$, but this can be generalized considerably. If we define
$\epsilon_t$ to be the noise in the gradient estimate at block $t$, \ie,
\[
\epsilon_t = \nabla f(p_t) -  \nabla_p \tilde f(p_t,\omega_t),
\]
then the above result requires that $\{ \epsilon_t \}$ be zero mean and i.i.d.
Under suitable conditions, this can be relaxed to the $\{\epsilon_t\}$ being a
martingale difference sequence~\cite[\S5.2]{harold1997stochastic}.

\subsection{Discussion}
With only the basic assumptions that (a) there is some bound on the maximum
price $p^\star$ that an adversary may choose and (b) some hard constraint on
the block size, which is enforced by the validators, we achieve average regret
that scales as $1/\sqrt{T}$ for either update rule, even when the transactions
and welfare, which are included in the functions $f_t$, are chosen
adversarially. An interesting point is that this shows why a hard constraint on
the resources consumed per-block is a good idea: it bounds any adversary's
ability to construct arbitrarily-sized blocks which could make the regret of
most reasonable rules quite large. The restriction given in (a), that the
adversary has bounded prices, seems strong, but is implied by a number of
simple statements: for example, it suffices that the adversary has finite
budget and must pay $p_t^TA_tx_t^\star(p_t)$ for each transaction, which is, in
fact, a much stronger requirement. Two interesting questions for future
research are: first, is there a more natural model for an adversary in the
blockchain setting than the essentially all-powerful one provided here? And,
second, given such an adversary, can we achieve better results?

\ifsubmit
    \subparagraph*{Time-independent step sizes.}
\else
    \paragraph{Time-independent step sizes.}
\fi
In this model, we assumed that the
step size $\eta$ is fixed and can depend on the total length of observed time
period, $T$. Indeed, the step size trades off the ability of the prices $p$ to
`react' to observed gradients $g_t$ with the average regret. (In particular, as
$T$ becomes large, the average regret becomes small, but $\eta$ must be
similarly small and must remain so for the time period in observation.) It is
not hard to show that the regret bound in both the adversarial
(\S\ref{sec:reg-bounds-update}) and stochastic case (\S\ref{sec:stochastic})
above for the gradient descent rule~\eqref{eq:grad-update} (\ie, the one
derived from the norm-square choice function) also carries through in the case
where the step sizes $\eta$ are decreasing with respect to time, to get an
update rule of the form
\begin{equation}\label{eq:update-time}
    x_{t+1} = x_t - \eta_tg_t.
\end{equation}
Here, $\eta_t = C/\sqrt{t}$ and $C> 0$ is some appropriately chosen constant
that, importantly, does not depend on the observation window $T$, but only on
the known bounds for the price $\|p\|_2 \le M$ and the maximum resource
utilizations $\|b\|_2 \le B$. The regret for this algorithm, under the same
assumptions as~\S\ref{sec:reg-bounds-update} for the norm-squared choice
function, is no more than $\sqrt{2}BM\sqrt{T}$ in the adversarial case;
see~\cite[\S3.1]{hazan2016introduction} for a proof. (The bound for the
stochastic case is more refined, but see, \eg,~\cite[Thm.\ 5.3]{garrigos2023handbook}
for the statement.) More generally, any sequence $\{\alpha_t\}$ which
asymptotically satisfies
\[
    \sum_{t=1}^T \alpha_t = O(\sqrt{T})
\]
and $1/\alpha_T = O(\sqrt{T})$ has a similar guarantee, up to a potentially
different constant.

\ifsubmit
    \subparagraph*{Strong convexity.}
\else
    \paragraph{Strong convexity.}
\fi
Indeed, if the step sizes of the algorithm are
allowed to vary with respect to time, we can get much stronger regret
guarantees in the case that the conjugate of the loss $\ell_t^\star$ is
strongly convex, using the same update rule~\eqref{eq:update-time}. More
specifically, if $\ell_t^\star$ is $\mu$-strongly convex, then we can achieve
\emph{logarithmic} regret that, additionally, does not depend on there being
any bound on the optimal price $p^\star$. (One case where $\ell_t^*$ is
strongly convex is, \eg, when $\ell_t(y) = (1/2)\|(y - b^\star)_+\|_2^2$ for
some target utilizations $b^\star \in \reals^m_+$ and $y$ is bounded by the
maximum resource utilizations $y \le b$.) In this particular case, we have that
for the update rule~\eqref{eq:update-time}, where the step sizes are $\eta_t =
\mu/t$, the regret is no larger than $B^2\log(T)/\mu$, making the average
regret quite small: no more than $B^2\log(T)/(\mu T)$. (For a proof of this,
see~\cite[\S3.3.1]{hazan2016introduction}.)

\section{Lower bound}
In this section, we will show that, up to a scalar constant, the bounds
provided in~\S\ref{sec:reg-bounds-update} are essentially optimal for any
algorithm. In particular, we will show that any rule to choose the prices
$p_{t+1}$, based on the previous $t$ observed gradients $g_1, \dots, g_t$,
achieves, at best, regret on the order of $1/\sqrt{T}$, for an adversary who
can control the transactions appearing in a block, subject to the constraints
of the set $S$ containing some maximum resource usage limit $b \in \reals_+^m$,
for a commonly-used loss. As in the previous section, we assume that the
optimal price $p^\star \in \reals^m$ for the aggregated program is bounded,
$\|p^\star\|_\infty \le M$ for some value $M > 0$.

\subsection{Construction}
The construction we provide here is very simple: the adversary is purely
stochastic and picks the resource utilizations, up to the bound, from a simple
distribution, independently of all of the previous gradients, such that
any algorithm achieves $0$ objective value, while the best possible price
$p$ achieves an objective of $-C\sqrt{T}$ for some constant $C > 0$.

\ifsubmit
    \subparagraph*{Loss function and set.}
\else
    \paragraph{Loss function and set.}
\fi
We construct a lower bound for any algorithm which
attempts to minimize the dual problem corresponding to the commonly-used loss
\begin{equation}\label{eq:loss-main}
    \ell(y) = \begin{cases}
        0 & 0 \le y \le b^\star\\
        +\infty & \text{otherwise},
    \end{cases}
\end{equation}
where $b^\star \in \reals_+^m$ is the resource utilization target. Using this definition,
it is not hard to show that the Fenchel conjugate of $\ell$ is
\[
    \ell^*(p) = (b^\star)^T(p)_+,
\]
where $((p)_+)_i = \max\{p_i, 0\}$ denotes the `positive part' of $p$, applied
elementwise. The set $S_t \subseteq \{0, 1\}^{n_t}$ we assume satisfies
\[
    S_t \subseteq \{x_t \in \reals_+^{n_t} \mid A_tx_t \le b\},
\]
for some matrix $A_t \in \reals^{m \times n_t}$ and max resource utilization $b
\in \reals_+^m$. (We note that this set up reduces exactly to the resource
metering of EIP-1559 when the number of resources $m=1$, where $n_t$ is the
number of transactions in the mempool of the block builder and $b$ is the block
limit.)

\ifsubmit
    \subparagraph*{Adversary.}
\else
    \paragraph{Adversary.}
\fi
Consider a (stochastic) adversary who is able to control
the amount of resources used by the network. The adversary picks the welfare to
be $q_t = 0$ and picks transactions $A_t$ such that the welfare maximization
problem~\eqref{eq:welfare-max} satisfies, at price $p_t$,
\[
    A_tx^\star_t = b^\star - \eps_t\circ \tilde b,
\]
where $\eps_t \sim \{\pm 1\}^m$ is uniformly randomly chosen (independently,
for each $t=1, \dots, T$) and
\[
    \tilde b_i = \min\{b^\star_i, b_i - b^\star_i\},
\]
for $i=1, \dots, m$. (The adversary can achieve this by constructing, say,
exactly one transaction that is always included with zero welfare and utilizations
equal to exactly $b^\star - \eps_t\circ \tilde b$; we assume
this in what follows.) This gives
\begin{equation}\label{eq:adv-grad}
    g_t = y^\star_t - A_tx_t^\star = b^\star - (b^\star - \eps_t\circ \tilde b) = \eps_t \circ \tilde b.
\end{equation}

\ifsubmit
    \subparagraph*{Proof.}
\else
    \paragraph{Proof.}
\fi
It suffices to show that, in expectation, the regret is
$\Omega(\sqrt{T})$, which means that there is at least one choice of resource
consumption trajectories (gradients $\{g_t\}$) that leads to large regret, for
any possible choices of prices $p_t$ by an algorithm.
First note that the linear bound~\eqref{eq:regret-bound},
\[
    \sum_{t=1}^T (f_t(p_t) - f_t(p^\star)) \le \sum_{t=1}^T g_t^T(p_t - p^\star),
\]
is met at exact equality for any nonnegative $p_t$ and $p^\star$ using our
construction. To see this, note that the transactions were chosen to have zero
welfare, $q_t = 0$. Using the definition of $\ell$~\eqref{eq:loss-main} means
that, for any $p, p_t \ge 0$:
\[
    f_t(p) = \ell^*_t(p) + h_t(p) = (b^\star)^T p + (q_t - A_t^Tx_t^\star(p_t))^Tp = (b^\star - A_tx_t^\star(p_t))^Tp = g_t^Tp,
\]
for every $t=1, \dots, T$. It will then suffice to show that the linear sum
\begin{equation}\label{eq:regret-bound-electric-boogaloo}
    \E\left[\sum_{t=1}^T g_t^T(p_t - p^\star)\right]
\end{equation}
is `large' in expectation, for $p^\star$ chosen with knowledge of $g_t$.

The first term of the loss~\eqref{eq:regret-bound-electric-boogaloo} is zero since
\[
    \E[g_t^Tp_t] = \E[g_t]^Tp_t = 0,
\]
as $p_t$ can depend only on the observed gradients, $g_1, \dots, g_{t-1}$.
The expected regret then reduces to
\[
    \E\left[\sum_{t=1}^T g_t^Tp_t - g_t^T\bar p\right] = \E\left[-\bar p^T\sum_{t=1}^T g_t\right].
\]
Since $p^\star$ is chosen after observing the gradients $g_t$, and we are free
to choose any $p^\star$ with $\|p^\star\|_\infty \le M$ and $p^\star \ge 0$, we
then this term is maximized by choosing $p^\star_i = M$ if
\[
    \sum_{t=1}^T (g_t)_i \le 0,
\]
and $0$ otherwise, for $i=1, \dots, m$. This means that
\[
    \E\left[-\bar p^T\sum_{t=1}^T g_t\right] = M\E\left[ \ones^T\left(-\sum_{t=1}^T g_t\right)_+\right],
\]
where $((z)_+)_i = \max\{z_i, 0\}$ is the `positive part' of $z_i$. Using the
definition of $g_t$ above~\eqref{eq:adv-grad}, we can bound the latter term
\[
    M\E\left[ \ones^T\left(-\sum_{t=1}^T g_t\right)_+\right] = M(\ones^T\tilde b)\E\left[\left(\sum_{t=1}^T \eps'_t\right)_+\right] \ge M(\ones^T\tilde b) C\sqrt{T},
\]
where $\eps'_t \sim \{\pm 1\}$ for $t=1, \dots, T$ is uniformly and
independently drawn and $C \ge 1/24$. The inequality follows from a mechanical
(if involved) counting argument, or any number of lower bounds on the expected
value of the positive side of a random walk. We reproduce a simple, noncounting
proof of this in appendix~\ref{app:noncounting-lower-bound}. We note that,
compared to either of the bounds of~\S\ref{sec:reg-bounds-update}, this
is tight up to a twiddle factor of $M$ and a square root factor of $m$.

\section{Conclusion}
In this paper, we have shown that two simple update rules: a gradient update
rule and a multiplicative update rule, lead to low average regret for the
discovery and assignment of prices for blockchain resources, even under
adversarial conditions. This implies that, on average, over a long enough
timescale, the difference between correctly setting resource prices with
oracular knowledge of the future (to roughly match supply and demand over the
period of time in question) and these two simple update rules, is very small.
We have also shown that, under the assumptions above, this is essentially the
best we can hope to do by also showing a matching lower bound. Since Ethereum's
EIP-1559 is the single-dimensional special case of the multiplicative rule
presented above, the analysis also applies directly, along with the proposed
pricing rule for EIP-4844, which is the two-dimensional special case. Future
potential avenues for research include weaker, but potentially more realistic
adversarial models (as the adversary has a lot of power in this particular
formulation) which may yield tighter bounds to the expected performance.

\ifsubmit
    \bibliographystyle{plainurl}% the mandatory bibstyle
\else
    \bibliographystyle{alpha}
\fi
\bibliography{citations.bib}

\appendix

\section{Alternative multiplicative choice function}\label{app:alt-mult-choice}
Instead of setting a bound on the gradients such that the multiplicative update
choice function of~\S\ref{sec:choice-functions} has arguments $z \in \reals^m$
with $\|z\|_\infty \le \log(M)$, we can instead choose the function $F$ in the
following way: if we know that $\|p\|_\infty \le M$ (or if we wish to have a
rule that satisfies this condition) then we can set the function $F$ to be,
instead,
\[
    F(z) = \sum_{i=1}^m \min\{(p_0)_i\exp(z_i), M(z_i - \log(M/(p_0)_i) + 1)\}.
\]
We then have that
\[
    (\nabla F(z))_i = \begin{cases}
        (p_0)_i\exp(z_i) & z_i \le \log(M/(p_0)_i)\\
        M & \text{otherwise}.
    \end{cases}
\]
It is not hard to show that $F$ is indeed convex by noting that it is both
separable and the derivative of any one term is nondecreasing by construction.
Additionally, $F$ is smooth with constant $mM$, which is easy to see by noting
that, again, $F$ is separable and the second derivative of each term is bounded
from above by $M$. By construction, we will have that any
\[
    \tilde p \in \nabla F(z),
\]
satisfies $\|\tilde p\|_\infty \le M$, as required, while
\[
    F^*(p) = \sum_{i=1}^m \left(p_i \log\left(\frac{p_i}{(p_0)_i}\right) - p_i\right),
\]
whenever $0 \le p \le M\ones$ (where we define $0\log 0 = 0$ for convenience).

\section{Noncounting lower bound}\label{app:noncounting-lower-bound}
In this appendix, we give a simple lower bound for the expected value of a
random walk. More specifically, given $\eps_t' \in \{\pm 1\}$ chosen uniformly
and independently for $t=1, \dots, T$, we will show that there exists a
constant $C' \ge 1/12$ such that
\[
    \E\left[\left|\sum_{t=1}^T \eps_t'\right|\right] \ge C'\sqrt{T}.
\]
Since $\eps_t'$ is symmetric about $0$ (\ie, $\eps_t'$ and $-\eps_t'$ have
the same distribution) then the positive part would satisfy
\[
    \E\left[\left(\sum_{t=1}^T \eps_t'\right)_+\right] \ge \frac{C'}{2}\sqrt{T} \ge \frac{1}{24} \sqrt{T},
\]
given the above. This material is a shorter version of~\cite{angeris2023noncounting}.

\ifsubmit
    \subparagraph*{Proof.}
\else
    \paragraph{Proof.}
\fi
For succinctness, we will write the sum as
\[
    X = \sum_{t=1}^T \eps_t'.
\]
The odd moments of $X$ vanish, while the second and fourth moments are
\[
    \E[X^2] = T, \quad \E[X^4] = 3T^2 - 2T \le 3T^2.
\]
which is easy to see from expanding the sum and noting that any term containing
an odd power of $\eps_t'$ has expectation equal to zero. The bound
will come from the simple observation that, for any $a \ge 0$,
\begin{equation}\label{eq:basic-pr-bound}
    \E[|X|] \ge a\Pr(|X| \ge a).
\end{equation}
We bound the second term in the inequality from below by expanding the second
moment,
\begin{equation}\label{eq:second-moment-bound}
    \E[X^2] \le a^2 + \E[X^2\ones_{|X| \ge a}] \le a^2 + \sqrt{\E[X^4]}\sqrt{\E[\ones_{|X| \ge a}]} = a^2 + \sqrt{\E[X^4]}\sqrt{\Pr(|X| \ge a)},
\end{equation}
for any fixed constant $a \ge 0$, where $\ones_{|X| \ge a}$ is the zero-one
indicator function for the event $|X| \ge a$. Here, the first inequality
follows from the pointwise inequality
\[
    X^2 \le a^2 \ones_{|X| < a} + X^2\ones_{|X| \ge a},
\]
while the second inequality follows from an application of Cauchy--Schwarz to
the latter term and the fact that $\ones_{|X| \ge a}^2 = \ones_{|X| \ge a}$.
Rearranging the second moment bound~\eqref{eq:second-moment-bound} above,
we find
\[
    \Pr(|X| \ge a) \ge \frac{(\E[X^2] - a^2)^2}{\E[X^4]} \ge \frac{(T - a^2)^2}{3T^2},
\]
where we have implicitly assumed that $a^2 \le T$. Plugging this in to the
basic bound~\eqref{eq:basic-pr-bound} on the expectation gives, for any $0 \le
a \le \sqrt{T}$,
\[
    \E[|X|] \ge a\frac{(T-a^2)^2}{3T^2}.
\]
The right hand quantity is maximized at $a=\sqrt{T/5}$ (which is easy to see by
using the first-order optimality conditions applied to $a$) so we get
\[
    \E[|X|] \ge \frac{16}{75\sqrt{5}}\sqrt{T} \ge \frac{1}{12}\sqrt{T},
\]
which means that $C' \ge 1/12$, as required. We note that this constant is
loose, as a simple counting argument shows that, in fact
\[
    C' \sim \sqrt{2/\pi},
\]
for $T$ large enough, but this bound is enough to show the lower bound holds up
to a constant.

\section{Follow-the-regularized-leader interpretation}\label{app:eip-1559}
We may interpret the price update rule~\eqref{eq:price-def} as
follow-the-regularized-leader (FTRL), with the regularizer defined by the choice
function. The FTRL update rule is given by
\[
    p_{t+1} = \argmin_{p \in D} \sum_{s=1}^t g_s^Tp + (1/\eta)R(p),
\]
where $R$ is some strongly convex regularizer and $D$ is the effective domain
of the $\{f_t\}$. From the optimality conditions, we can write
\[
    p_{t+1} = (\nabla R)^{-1}(-\eta(g_1 + \dots + g_t)),
\]
assuming the range of $(\nabla R)^{-1}$ is contained in the domain $D$. (Note
that $\nabla R: D \to \reals^n$ is an invertible function since $R$ is
strongly convex.) Comparing this equation with the choice function-based price
update,
\[
    p_{t+1} = \nabla F(-\eta(g_1 + \dots + g_t)),
\]
we immediately see the relationship between the choice function $F$ and the
regularizer $R$, namely that $\nabla R = (\nabla F)^{-1}$ and vice versa.
(Equivalently, this is saying that $R$ and $F$ are Fenchel conjugates of one
another.) We may alternatively interpret the update rule as online mirror
descent with Bregman divergence defined by the regularizer.
See~\cite{mcmahan2011follow} for details.

\ifsubmit
    \subparagraph*{EIP 1559-style price updates.}
\else
    \paragraph{EIP 1559-style price updates.}
\fi
This connection allows us to interpret the multiplicative update
rule~\eqref{eq:mult-update} as FTRL. Consider the loss function $\ell_t(y) = I(y
= b^\star)$ for some target gas usage $b^\star \in \reals_+$. The matrix $A_t$ is simply a
row vector of the gas costs for each transaction. The gradient at block $t$,
which has gas price $p_t$, is then
\[
    g_t = \nabla f_t(p_t) = b^\star - A_tx_t^\star(p_t).
\]
From section~\ref{sec:choice-functions}, we write the EIP-1559-style 
multiplicative update rule with $p_0 = 1$ as
\[
    p_{t+1} = p_t \circ \exp(-\eta g_t) = \nabla F(-\eta(g_1 + \dots + g_t)).
\]
As discussed above, we may also interpret this update rule as FTRL with
the regularizer
\[
    R(p) = p\log(p) - p.
\]
This regularizer achieves its minimum at $p = p_0 = 1$. Note that it is not
symmetric around its minimum; price decreases are more quickly penalized than
price increases.

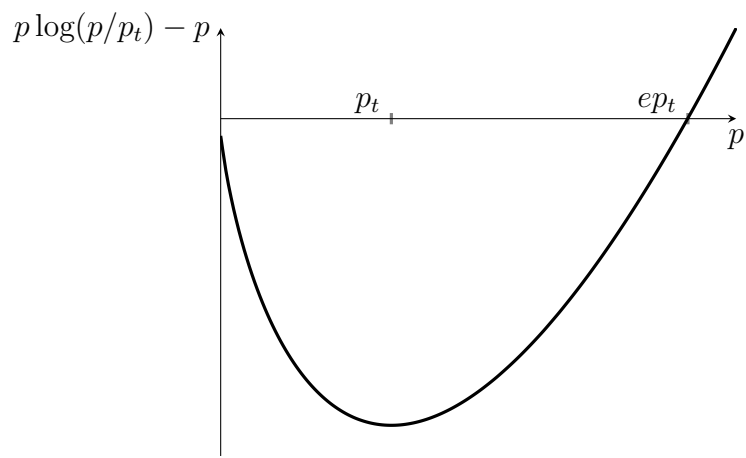
\begin{figure}
    \centering
    \begin{tikzpicture}
    \begin{axis}[
        xlabel={$p$},
        ylabel={$p \log(p/p_t) - p$},
        domain=0.01:3,
        samples=100,
        smooth,
        no markers,
        axis lines=center,
        x label style={anchor=north},
        y label style={anchor=east},
        xtick={1, 2.72},
        xticklabels={$p_t$, $ep_t$},
        xticklabel style={above left},
        xtick style={very thick},
        ymin=-1.1,
        ytick=\empty
    ]
    \addplot[black, very thick, domain=0.01:3.00]{x * ln(x) - x};
    \end{axis}
    \end{tikzpicture}
    \caption{Bregman divergence corresponding to the nonnegative entropy 
    function, which is used in the EIP 1559-style update rule}
    \label{fig:bregman}
\end{figure}

\ifsubmit
    \subparagraph*{EIP 1559 as online mirror descent.}
\else
    \paragraph{EIP 1559 as online mirror descent.}
\fi
Alternatively, we may interpret the EIP-1559-style multiplicative price update
rule as online mirror descent with Bregman divergence defined by the nonnegative
entropy $p\log(p)$. This update rule is given by
\[
\begin{aligned}
    p_{t+1} = \argmin_{p \in \dom f} f(p_t) + g_t^T(p - p_t) + \frac{1}{\eta}D(p,\, p_t),
\end{aligned}
\]
where the first two terms are a linear approximation to the convex function $f$
and the last term is the Bregman divergence regularizer
\[
    D(p,\, p_t) = p\log(p) - p_t \log(p_t) - (\log(p_t) + 1)(p - p_t).
\]
Dropping constant terms, we can rewrite the update rule as
\[
    \begin{aligned}
        p_{t+1} &= \argmin_{p \in \dom f} g_t^Tp + \frac{1}{\eta}\left(p\log(p / p_t) - p \right) \\
        &= p_t \circ \exp(-\eta g_t).
    \end{aligned}
\]
Note that the regularizer $D(p,\, p_t)$ achieves its minimum at $p = p_t$ and
penalizes price decreases more than price increases. (See
figure~\ref{fig:bregman}.) This asymmetric penalty may explain some past
observations, \eg, that EIP 1559 has a positive
overshoot~\cite{leonardos2023optimality}.

\end{document}